\documentclass{article}
\usepackage{amsfonts}
\usepackage{amsmath}
\usepackage{amssymb}
\usepackage{graphicx}
\usepackage{array}
\usepackage{booktabs}
\usepackage{multirow}
\usepackage{makecell}
\usepackage{float}
\usepackage{booktabs}
\pdfoutput=1 

\usepackage{mymacros}
\usepackage{chngcntr}
\usepackage{verbatim}
\usepackage{amsthm}
\usepackage{graphics}
\usepackage{xcolor}
\usepackage{mathrsfs}
\usepackage{bbold}
\usepackage{cases}
\usepackage{epsfig}
\usepackage{epstopdf}
\usepackage{hyperref}
\usepackage{amsfonts}
\usepackage{hyperref}
\usepackage{jheppub} 

\usepackage[T1]{fontenc} 

\hypersetup{hypertex=true,
	colorlinks=true,
	linkcolor=blue,
	anchorcolor=blue,
	citecolor=blue}
\usepackage[numbers,sort&compress]{natbib}

\title{Topology of Ho\v{r}ava-Lifshitz black holes in different ensembles}
\author[a]{Deyou Chen, \footnote{E-mail: deyouchen@hotmail.com }}
\author[a]{Yucheng He, \footnote{E-mail: heyucheng365@hotmail.com }}
\author[b]{Jun Tao, \footnote{E-mail: taojun@scu.edu.cn }}
\author[c]{Wei Yang \footnote{E-mail: yangweicdu@hotmail.com }}

\affiliation{$^{a}$School of Science, Xihua University, Chengdu 610039, China}
\affiliation{$^{b}$Center for Theoretical Physics, College of Physics, Sichuan University, Chengdu 610065, China}
\affiliation{$^{c}$School of Information Science and Engineering, Chengdu University, Chengdu 610106, China}

\abstract{In this paper, we study topological numbers for uncharged and charged static black holes obtained in Ho\v{r}ava-Lifshitz gravity theory in different ensembles. We first calculate the topological numbers for the uncharged black holes by changing the value of the dynamic coupling constant, and find that the black holes with spherical and flat horizons have the same topological number. When the black hole's horizon is hyperbolic, different values of the coupling constant generate different topological numbers, which can be $1$, $0$ or $-1$. This shows that the coupling constant plays an important role in the topological classification. Then, we study the topological numbers for the charged black holes in different ensembles. The black hole with a spherical horizon has the same topological number in canonical and grand canonical ensembles. When the horizons are flat or hyperbolic, they have different topological numbers in canonical and grand canonical ensembles. Therefore, the topological numbers for the uncharged black holes are parameter dependent, and those for the charged black holes are ensemble dependent. }

\begin{document}
	\maketitle

\section{Introduction}

Topological approaches are effective ways for studying physical systems, which ignore the specific structure of the systems and focus on their general characteristics. Defects are important tools for these approaches, as they reveal certain properties of field configurations. Recently, a topological approach has been used to research on the black holes \cite{WLM}. In this elegant approach, the black hole solutions are treated as defects in the thermodynamic parameter space. Local thermodynamic stability and instability of a black hole are determined by positive and negative winding numbers, respectively. The global characteristic is characterized by a topological number which is sum of the winding numbers for all the black hole branches at an arbitrary temperature. Each black hole is endowed with a topological number, and then black holes can be classified according to the values of the numbers. This approach was based on Duan's $\phi$-mapping topological current theory \cite{DL1,DL2}, and a key point is the construction of a vector field. In \cite{WLM}, they defined the field through the introduction of a generalized free energy. Based on this work, the influences of the cosmological constant, angular momentum parameters, charges, and other parameters on the topological numbers in the different spacetimes have been extensively researched, and many important results have been found \cite{BST,YBM1,YBM2,YBM3,YBM4,WU1,WU2,WU3,WU4,WU5,WU6,LW1,LW2,FJZ,DZ1,DZ2,ZJ,FS1,FS2,FS3,FS4,FS5,FS6,FS7,FS8,AAGS,GP1,GP2,GP3,ZCH,CHT,AMKM}. In black hole physics, topological approaches are not limited to the study of black hole classification, but are also used to study light rings and shadows, as well as other natures of black holes \cite{CBH,GG1,GG2,SWW1,SWW2,WL1,WL2}.

In this paper, we study topological numbers for uncharged and charged black holes obtained in the theory of Ho\v{r}ava-Lifshitz(HL) gravity in different ensembles. The influence of a dynamical coupling constant for the uncharged black holes and that of different ensembles for the charged black holes on the topological numbers are discussed. HL theory was proposed by Ho\v{r}ava \cite{HL1,HL2,HL3}, which is a non-relativistic renormalizable theory of gravity at a Lifshitz point. At short distances, this theory describes interacting nonrelativistic gravitons. When the condition of detailed balance is restrictively obeyed, it is intimately related to topologically massive gravity in three dimensions. At long distances, it reduces to the relativistic value $z = 1$, where $z$ measures the degree of anisotropy between space and time. Therefore, this theory is seen as a candidate for the UV completion of Einstein's general relativity. Since the theory was proposed, it immediately attracted much attention. L\"u first obtained the solutions for spherically symmetric black holes and Friedman-Lema\^itre-Robertson-Walker cosmology from the HL gravity action \cite{LMP}. Subsequently, Cai et al considered a general dynamical coupling constant, and obtained the solution of the topological black holes and discussed their thermodynamic properties \cite{CCO1,CCO2,CCS}. Compared with other UV complete gravity theories, HL theory exhibits significantly different UV behaviors. Meanwhile, different values of the coupling constant and different ensembles may affect topological numbers. Therefore, it is necessary to study the topological properties of this theory and the influence of the dynamic coupling constant on topological numbers.

The rest is organized as follows. In the next section, we give a brief review of the topological approach propsoed in \cite{WLM}. In Section 3, we study the influence of different values of the dynamic coupling constant on the topological numbers for the uncharged black holes. In Section 4, the influence of different ensembles on the topological numbers for the charged black holes is studied. The last section is devoted to our conclusion and discussion.

\section{Review of topological approach}

In \cite{WLM}, the generalized free energy is defined by

\begin{eqnarray}
\mathcal{F} = E- \frac{S}{\tau},
\label{3.1.1}
\end{eqnarray}

\noindent where $E$ and $S$ are the energy and entropy of a system, respectively. $\tau$ is a variable and can be seen as the inverse temperature of the cavity enclosing the black hole. This free energy is off-shell except at $\tau = 1/T$. An important vector is constructed via a thermodynamic approach,

\begin{eqnarray}
\phi = \left(\frac{\partial \mathcal{F}}{\partial r_h} , -\cot\Theta \csc\Theta\right).
\label{3.1.2}
\end{eqnarray}

\noindent where $0<r_h<+\infty$ and $0\le\Theta\le\pi$. Zero points of the vector obtained at $\tau = 1/T$ and $\Theta = \pi/2$ correspond to the on-shell black hole solution. Other points are not the solutions of Einstein field equations, and then they are the off-shell states. $\phi^{\Theta}$ diverges at $\Theta =0$ and $\Theta =\pi$, which leads to that the direction of the vector is outward.

Using Duan's $\phi$-mapping topological current theory, one can define a topological current \cite{DL1,DL2}

\begin{eqnarray}
j^{\mu} = \frac{1}{2\pi}\varepsilon^{\mu\nu\rho}\varepsilon_{ab}\partial_{\nu}n^a\partial_{\rho}n^b,
\label{3.1.3}
\end{eqnarray}

\noindent where $\mu,\nu,\rho = 0,1,2$, $a,b = 1,2$, $\partial_{\nu} = \frac{\partial}{x^{\nu}}$ and $x^{\nu} = (\tau, r_h, \Theta)$. $\tau$ is seen as a time parameter of the topological defect, and $n^a$ is a unit vector defined by $\left( \frac{n^r}{||n||}, \frac{n^{\Theta}}{||n||}\right)$. It is easy to prove that current is conserved. Using the Jacobi tensor $\varepsilon^{ab}J^{\mu}(\frac{\phi}{x}) = \varepsilon^{\mu\nu\rho} \partial_{\nu}{\phi}^a\partial_{\rho}{\phi}^b$ and two-dimensional Laplacian Green function $\Delta_{\phi^a}ln||\phi||=2\pi \delta^2(\phi)$, the current is rewritten as

\begin{eqnarray}
j^{\mu} = \delta^2(\phi)j^{\mu}\left(\frac{\phi}{x}\right),
\label{3.1.4}
\end{eqnarray}

\noindent which is nonzero only when $\phi^a(x^i) = 0$. Then a topological number in a parameter region $\sum$ is obtained as follows

\begin{eqnarray}
W= \int_{\sum}j^{0}d^2x= \sum_{i=1}^{N}\beta_i\eta_i = \sum_{i=1}^{N}w_i,
\label{3.1.5}
\end{eqnarray}

\noindent where $j^{0} = \sum_{i=1}^{N}\beta_i\eta_i\delta^2(\vec{x}-\vec{z}_i)$ is the density of the current, $\beta_i$ is Hopf index which counts the number of the loops that $\phi^a$ makes in the vector $\phi$ space when $x^{\mu}$ goes around the zero point $z_i$. Clearly, this index is always positive. $\eta_i$=sign$J^0(\phi/x)_{z_i} = \pm 1$ is the Brouwer degree. $w_i$ is the winding number for the $i$-th zero point of the vector in the region and its values is independent on the shape of the region.

\section{Topological numbers for uncharged HL black holes}

The action of Horava-Lifshitz theory is written as \cite{HL1,HL2}

\begin{eqnarray}
I &=& \int{dtd^3x(\mathcal{L}_0 + \mathcal{L}_1)}, \nonumber\\
\mathcal{L}_0 &=& \sqrt{g}N \left[\frac{2}{\kappa^2}(K_{ij}K^{ij}-\lambda K^2) + \frac{\kappa^2\mu^2(\Lambda R -3\Lambda^2)}{8(1-3\lambda)}\right],\nonumber\\
\mathcal{L}_1 &=& \sqrt{g}N \left[\frac{\kappa^2\mu^2(1 -4\lambda)}{32(1-3\lambda)}R^2 -\frac{\kappa^2}{2\omega^4}\left(C_{ij} - \frac{\mu \omega^2 R_{ij}}{2}\right)\left(C^{ij} - \frac{\mu \omega^2 R^{ij}}{2}\right)\right],
\label{eq2.1}
\end{eqnarray}

\noindent where $\kappa^2$, $\lambda$, $\mu$, $\omega$ and $\Lambda$ are constant parameters, and the Cotten tensor, $C_{ij}$, is defined by

\begin{eqnarray}
C_{ij} &=& \epsilon^{ikl}\Delta_k (R^i_l- \frac{1}{4}R \delta^i_l).
\label{eq2.2}
\end{eqnarray}

\noindent Comparing the action to that of general relativity, one can get the speed of light, Newton's constant and the cosmological constant,

\begin{eqnarray}
c=\frac{\kappa^2 \mu}{4}\sqrt{\frac{\Lambda}{1- 3\lambda}}, \quad G=\frac{\kappa^2}{32\pi c}, \quad \Lambda_C = \frac{3}{2}\Lambda.
\label{eq2.3}
\end{eqnarray}

\noindent $\mathcal{L}_0$ is equivalent to the usual Einstein-Hilbert Lagrangian when $\lambda = 1$. In the HL theory, $\lambda$ is the dynamical coupling constant, susceptible to quantum corrections \cite{HL1}. The cosmological constant is negative when $\lambda > 1/3$, and is positive when $\lambda < 1/3$. In this paper, we only consider the negative cosmological constant.

Cai et al considered a general dynamical coupling constant $\lambda$, and obtained the solution of the topological black holes in the HL gravity \cite{CCO1,CCO2}. The metric is

\begin{eqnarray}
ds^2=-N^2(r)f(r)dt^2+\frac{dr^2}{f(r)}+ r^2 d\Omega_{k}^2,	
\label{2.4}	
\end{eqnarray}	

\noindent where $d\Omega_{k}^2$ is the line element for two-dimensional space with constant scalar curvature $2k$. Without loss of generality, we take $k = 1, 0$ and $ -1$ which implies the spherical, flat and hyperbolic horizons, respectively. The functions $f(r)$ and $N(r)$ are given by

\begin{eqnarray}
f(r)=k-\Lambda r^2-\alpha r^s, \quad N(r)=\gamma r^{1-2s},
\label{2.5}	
\end{eqnarray}	

\noindent where $\alpha$ and $\gamma$ are both integration constants, and $s$ is defined by $s=\frac{2\lambda\pm \sqrt{2(3\lambda-1)}}{\lambda-1}$. In \cite{CCS}, the authors studied the thermodynamical properties of the charged and uncharged topological black holes, and found some interesting result which were never observed in Einstein gravity. The cosmological constant was seen as a fixed constant in the past. Recently, it has been regarded as a variable related to pressure, $P = -\frac{\Lambda}{8\pi} = \frac{3}{8\pi l^2}$, and its conjugate quantity is a thermodynamic volume $V$. In this paper, we use the initial expression of the cosmological constant. The mass, Hawking temperature, entropy are \cite{CCO1}

\begin{eqnarray}
	M&&=\frac{\sqrt{2} \kappa^2 \mu^2\Omega_{k} \gamma}{16\sqrt{3\lambda-1}}\frac{(k-\Lambda r_h)^2}{r_h^{2s}},\nonumber\\
    T&&=\frac{\gamma}{4\pi r_h^{2s}}[-\Lambda r_h^2(2-s)-ks], \nonumber\\
	S&&=\frac{\pi \kappa^2 \mu^2\Omega_{k}}{\sqrt{2(3\lambda-1)}}[k \ln(\sqrt{-\Lambda }r_h)+\frac{1}{2}(\sqrt{-\Lambda}r_h)^2]+S_0.
\label{2.6}	
\end{eqnarray}	

\noindent In the above equations, $r_h$ is the horizon radius and $S_0$ is a constant. When the speed of light $c=\frac{\kappa^2 \mu}{4}\sqrt{\frac{\Lambda}{1-3\lambda}} = \frac{2-s}{1+s}\frac{\kappa^2 \mu \sqrt{-\Lambda}}{4\sqrt{2}}$ and Newton's constant are adopted, the mass and entropy are rewritten as follows \cite{CCS}

\begin{eqnarray}
	M&&=\frac{c^3\gamma \Omega_k\Lambda}{16\pi G}\frac{1+s}{s-2}\left(k- \Lambda r_h^2\right)^2r_h^{-2s}, \nonumber\\
    S&&=\frac{c^3\Omega_k\Lambda}{4G}\frac{1+s}{s-2}[k \ln(-\Lambda r_h^2 )-\Lambda r_h^2 ]+S_0.
\label{2.7}	
\end{eqnarray}	

To study the topological properties, we adopt the definition of the generalized free energy and get

\begin{eqnarray}
	\mathcal{F}=\frac{c^3\gamma \Omega_k\Lambda}{16\pi Gr_h^{2s}}\frac{1+s}{s-2}\left(k- \Lambda r_h^2\right)^2-\frac{c^3\Omega_k\Lambda[k \ln(-\Lambda r_h^2 )-\Lambda r_h^2 ]\frac{1+s}{s-2}+4G S_0}{4G\tau}.
\label{2.8}	
\end{eqnarray}	

\noindent We calculate the vector $\phi$ and obtain its components,

\begin{eqnarray}
	\Phi^{r_h} &=& \frac{c^3\gamma \Omega_k\Lambda\left(k- \Lambda r_h^2\right)}{8\pi Gr_h^{2s+1}}\frac{1+s}{2-s}\left[2\Lambda +s\left(k- \Lambda r_h^2\right)\right] -\frac{c^3\Omega_k\Lambda}{4G\tau r_h}\frac{1+s}{s-2}\left(k- \Lambda r_h^2\right),\nonumber\\
	\Phi ^{\Theta} &=& -\cot\Theta \csc\Theta.
	\label{2.9}
\end{eqnarray}	
	
\begin{figure}[H]
	\centering
	\begin{minipage}[t]{0.48\textwidth}
		\centering
		\includegraphics[width=0.6\linewidth]{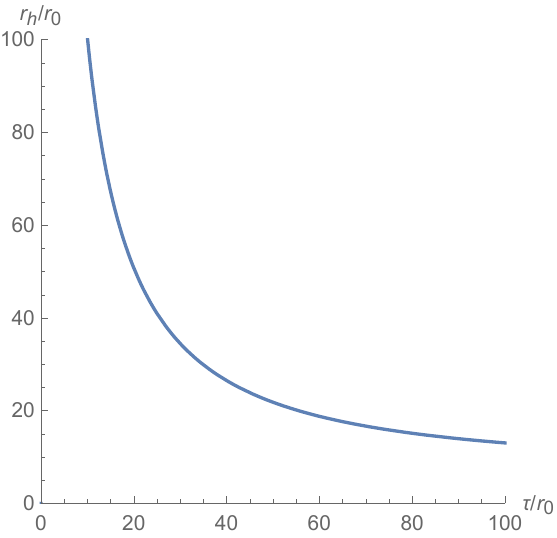}
	\end{minipage}
	\begin{minipage}[t]{0.3\textwidth}
		\centering
		\includegraphics[width=1\linewidth]{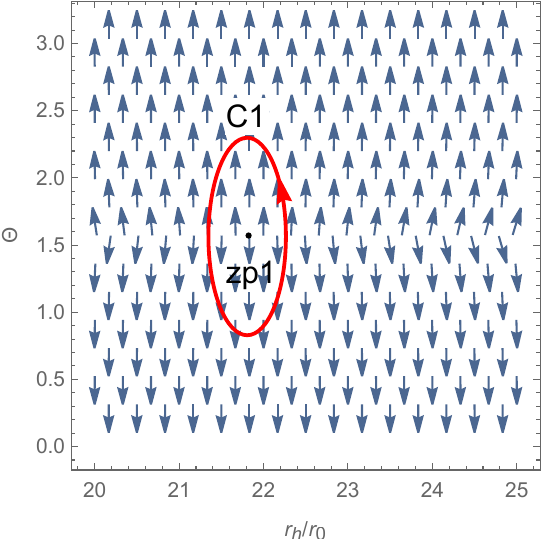}
	\end{minipage}
	\caption{Topological properties of the uncharged HL black hole, where $k = 1$, $s = 1/2$, $\lambda = 1$, $\gamma = 1$ and $\Lambda r_0^2 = -0.00838$. Zero points of the vector $\phi^{r_h}$ in the plane $r_h - \tau$ are plotted in the left picture. The unit vector field $n$ on a portion of the plane $\Theta - r_h $ at $\tau /r_0=50.00 $ is plotted in the right picture. The zero point is at $(r_h/r_0 ,\Theta)$= ($21.82, \pi/2$).}		
	\label{fig:2.1.1}
\end{figure}

\begin{figure}[H]
	\centering
	\begin{minipage}[t]{0.48\textwidth}
		\centering
		\includegraphics[width=0.6\linewidth]{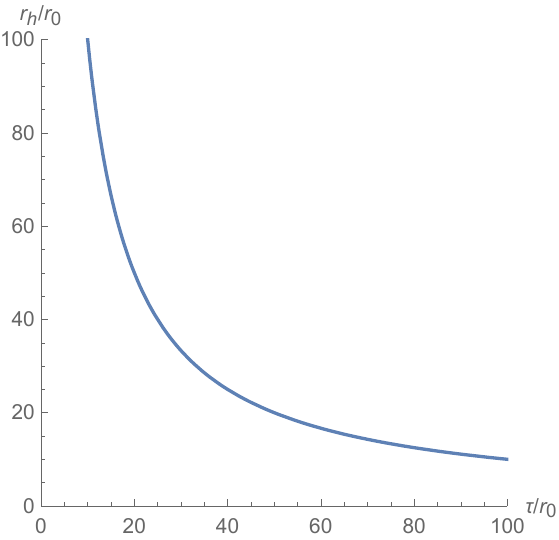}
	\end{minipage}
	\begin{minipage}[t]{0.3\textwidth}
		\centering
		\includegraphics[width=1\linewidth]{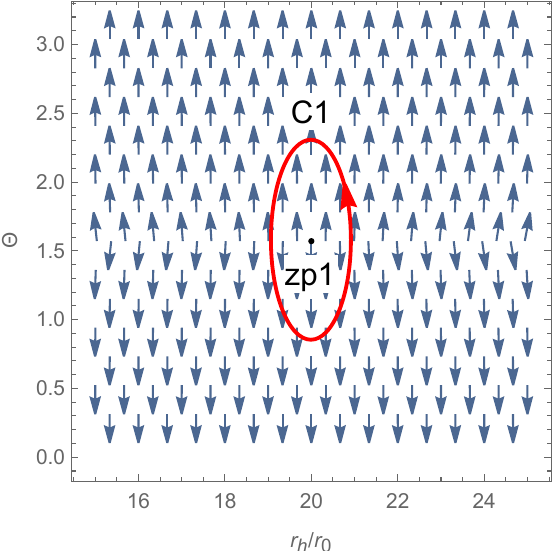}
	\end{minipage}
\caption{Topological properties of the uncharged HL black hole, where $k = 0$, $s = 1/2$, $\lambda = 1$, $\gamma = 1$ and $\Lambda r_0^2 = -0.00838$. Zero points of the vector $\phi^{r_h}$ in the plane $r_h - \tau$ are plotted in the left picture. The unit vector field $n$ on a portion of the plane $\Theta - r_h $ at $\tau /r_0=50.00 $ is plotted in the right picture. The zero point is at $(r_h/r_0 ,\Theta)$= ($20.00, \pi/2$).}			
\label{fig:2.1.2}
\end{figure}

\begin{figure}[H]
	\centering
	\begin{minipage}[t]{0.48\textwidth}
		\centering
		\includegraphics[width=0.6\linewidth]{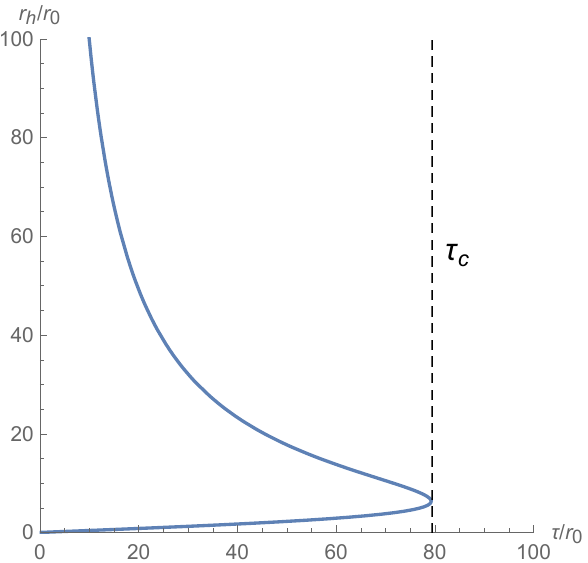}
	\end{minipage}
	\begin{minipage}[t]{0.3\textwidth}
		\centering
		\includegraphics[width=1\linewidth]{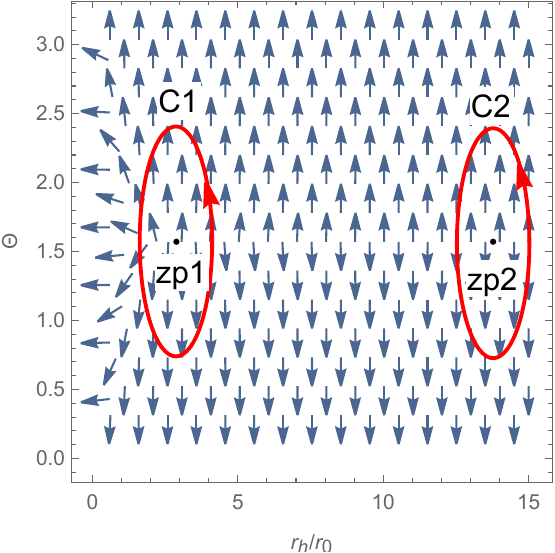}
	\end{minipage}
\caption{Topological properties of the uncharged HL black hole, where $k = -1$, $s = 1/2$, $\lambda = 1$, $\gamma = 1$ and $\Lambda r_0^2 = -0.00838$. Zero points of the vector $\phi^{r_h}$ in the plane $r_h - \tau$ are plotted in the left picture. The unit vector field $n$ on a portion of the plane $\Theta - r_h $ at $\tau /r_0=60.00 $ is plotted in the right picture. Zero points are at $(r_h/r_0 ,\Theta)$= ($2.89, \pi/2$) and ($13.78, \pi/2$), respectively.}			
\label{fig:2.1.3}
\end{figure}

\begin{figure}[H]
	\centering
	\begin{minipage}[t]{0.48\textwidth}
		\centering
		\includegraphics[width=0.6\linewidth]{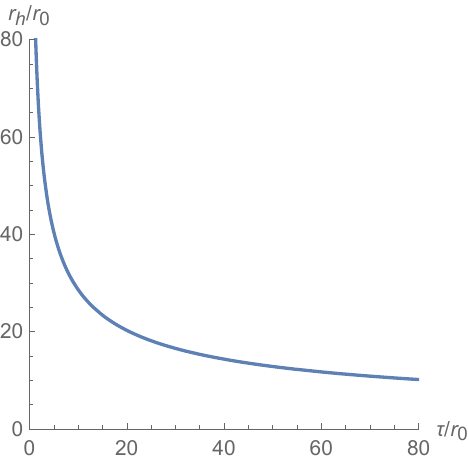}
	\end{minipage}
	\begin{minipage}[t]{0.3\textwidth}
		\centering
		\includegraphics[width=1\linewidth]{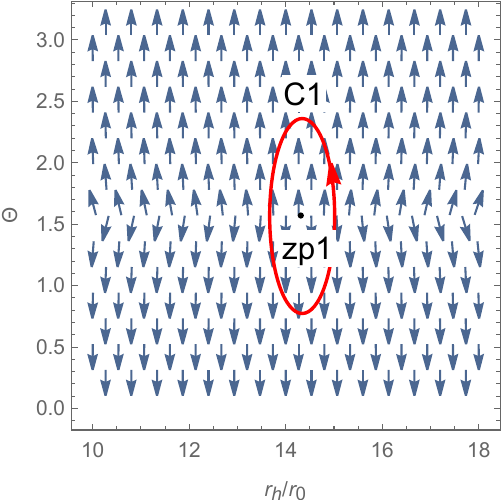}
	\end{minipage}
\caption{Topological properties of uncharged HL black holes, where $k = 1,0$ or $-1$, $s =0$, $\lambda = 1/2$ and $\Lambda r_0^2 = -0.00838$. Zero points of the vector $\phi^{r_h}$ in the plane $r_h - \tau$ are plotted in the left picture. The unit vector field $n$ on a portion of the plane $\Theta - r_h $ at $\tau /r_0=40.00 $ is plotted in the right picture. The zero point is at $(r_h/r_0 ,\Theta)$= ($14.31, \pi/2$).}			
\label{fig:2.1.4}
\end{figure}

\begin{figure}[H]
	\centering
	\begin{minipage}[t]{0.48\textwidth}
		\centering
		\includegraphics[width=0.6\linewidth]{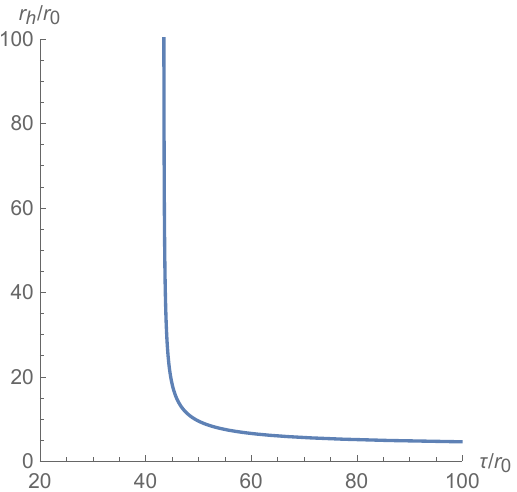}
	\end{minipage}
	\begin{minipage}[t]{0.3\textwidth}
		\centering
		\includegraphics[width=1\linewidth]{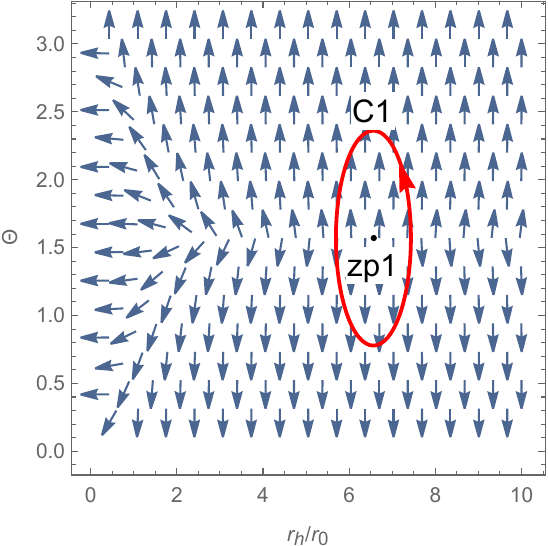}
	\end{minipage}
\caption{Topological properties of the uncharged HL black hole, where $k = 1$, $s =1$, $\lambda = 3$ and $\Lambda r_0^2 = -0.08380$. Zero points of the vector $\phi^{r_h}$ in the plane $r_h - \tau$ are plotted in the left picture. The unit vector field $n$ on a portion of the plane $\Theta - r_h $ at $\tau /r_0=60.00 $ is plotted in the right picture. The zero point is at $(r_h/r_0 ,\Theta)$= ($6.57, \pi/2$).}			
\label{fig:2.1.5}
\end{figure}

\begin{figure}[H]
	\centering
	\begin{minipage}[t]{0.48\textwidth}
		\centering
		\includegraphics[width=0.6\linewidth]{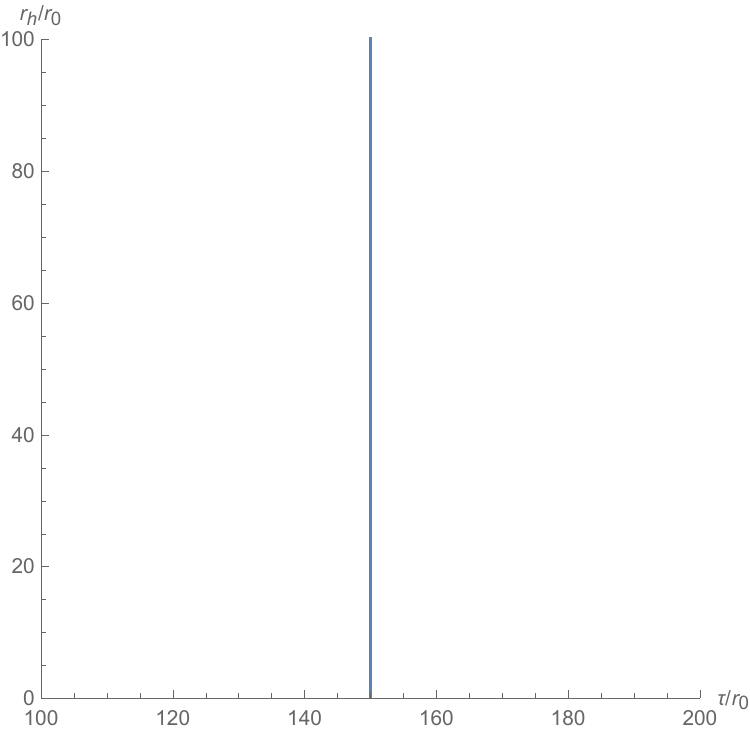}
	\end{minipage}
\caption{Zero points of the vector $\phi^{r_h}$ in the plane $r_h - \tau$, where $k = 0$, $s =1$, $\lambda = 3$ and $\Lambda r_0^2 =- 0.08380$.}			
\label{fig:2.1.6}
\end{figure}

\begin{figure}[H]
	\centering
	\begin{minipage}[t]{0.48\textwidth}
		\centering
		\includegraphics[width=0.6\linewidth]{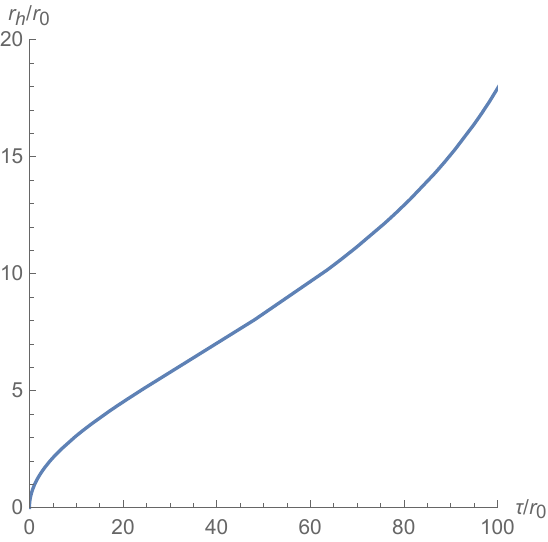}
	\end{minipage}
	\begin{minipage}[t]{0.3\textwidth}
		\centering
		\includegraphics[width=1\linewidth]{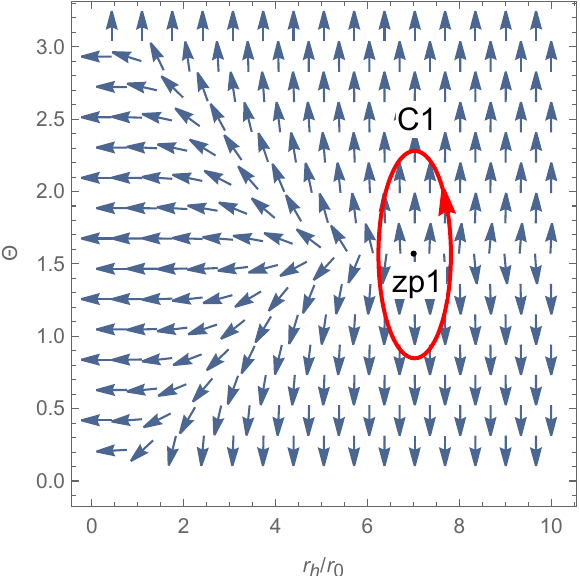}
	\end{minipage}
\caption{Topological properties of the uncharged HL black hole, where $k = -1$, $s =1$, $\lambda = 3$ and $\Lambda r_0^2 = -0.00838$. Zero points of the vector $\phi^{r_h}$ in the plane $r_h - \tau$ are plotted in the left picture. The unit vector field $n$ on a portion of the plane $\Theta - r_h $ at $\tau /r_0=40.00 $ is plotted in the right picture. The zero point is at $(r_h/r_0 ,\Theta)$= ($7.01, \pi/2$).}			
\label{fig:2.1.7}
\end{figure}

\noindent Zero points of $\phi^{r_h}$ determine the topological properties. Let $\phi^{r_h} = 0$ and get the relation between $r_h$ and $\tau$,

\begin{eqnarray}
		\tau =-\frac{2 \pi r_h^{2 s}}{\gamma  \left[-2\Lambda r_h^2 -s(k- \Lambda r_h^2)\right]},
\label{2.10}	
\end{eqnarray}	

\noindent which shows the change of the inverse temperature with the horizon radius. To clearly and intuitively study the local and global properties of the black holes, we use Eqs. (\ref{2.9}) and (\ref{2.10}) and plot Figure (\ref{fig:2.1.1}) - (\ref{fig:2.1.7}). The influence of the dynamical coupling constant $\lambda$ on the topological numbers is inverseed in these figures. During the calculation, $c= G=\Omega_k =1$, and $r_0$ is the length scale of the cavity surrounding the black holes.

In Figure (\ref{fig:2.1.1}), the horizon radius decreases monotonically with the increase of $\tau$'s value, which shows that the black hole is stable for any temperature and its winding number is 1. We order $\tau /r_0=50.00 $ and plot the picture of the unit vector field in the right picture of the figure. There is one zero point which is at $(r_h/r_0 ,\Theta)$= ($21.82, \pi/2$) in the picture. This zero point yields the winding number of $1$. Thus the topological number for this black hole is 1. In \cite{WU2}, the authors found that the topological number for a four-dimensional Schwarzschild AdS black hole is 0. Therefore, this black hole is different from the Schwarzschild AdS black hole in topological class.

When $k=0$, the horizon is flat and the topological properties are reflected in Figure (\ref{fig:2.1.2}). It is clearly from the left picture that the topological number is also 1. And then the winding number yielded by the zero point is $1$.

When $k=-1$, the metric (\ref{2.4}) denotes the black hole with a hyperbolic horizon and its topological properties are shown in Figure (\ref{fig:2.1.3}). In the left picture of the figure, an annihilation point divides the black hole into a stable region and an unstable region, which yield the winding numbers are $1$ and $-1$, respectively. There are two horizon radii for a same $\tau$ when $\tau < \tau_c$, and they coincide with each other at the annihilation point. When $\tau > \tau_c$, there is no black hole existed in the left picture. The positions of the zero points are shown in the right picture of the figure and are located at ($2.89, \pi/2$) and ($13.78, \pi/2$), respectively. The winding numbers corresponding to these two points are 1 and -1, respectively. Thus the topological number is $0$.

When $\lambda=1/2 $, there is $s=0 $, and it is easy to find from Eq. (\ref{2.10}) that the value of $ \tau $ is independent of $k $. Thus the topological properties of the black holes with the spherical, flat and hyperbolic horizons are shared by Figure (\ref{fig:2.1.4}). In the figure, the horizon radii decrease monotonically with the increase of $\tau$'s value, which shows that the black holes are stable for any temperature and the winding number is 1. The black holes have a same topological number $1$. The zero points are located at the same position. This shows that the dynamical coupling constant $\lambda$ plays an important role in the topological class of the black holes.

When $\lambda=3$, we get $s=1$ and plot Figure (\ref{fig:2.1.5}) - (\ref{fig:2.1.7}). From Figure (\ref{fig:2.1.5}), we find that when the $\tau$'s value decreases at a certain value, the horizon radius increase very fast. The black hole is stable for any $\tau$'s value and its winding number is 1. The zero point is at $(r_h/r_0 ,\Theta)$= ($6.57, \pi/2$). Therefore, its topological number is 1.

When $k=0$, it is easy to find from Eq. (\ref{2.10}) that the $\tau$'s value is independent on the horizon radius. Therefore, the variation of $r/r_0$ with $\tau /r_0$ is a vertical line parallel to the vertical axis in Figure (\ref{fig:2.1.6}). The horizon radius is not affected by the $\tau$'s value, and the black hole is stable. It is natural to obtain the topological number of 1.

When $k=-1$, Figure (\ref{fig:2.1.7}) is plotted. In the figure, the horizon radius increase monotonically with the $\tau$'s value, which shows that the black hole is unstable for any $\tau$'s value and its winding number is $-1$. The zero point is at ($7.01, \pi/2$) for $\tau /r_0=40.00 $ and corresponds to the winding number $-1$. Therefore, the topological number is $-1$.

Comparing Figure (\ref{fig:2.1.1}) - (\ref{fig:2.1.7}), it is evident that the uncharged black hole with the spherical and flat horizons have the same topological number. However, for the different values of $\lambda$ in the uncharged black hole with the hyperbolic horizon, its topological numbers can be $1$, $ 0$ or $-1$. Therefore, its topological number is parameter dependent.

\section{Topological numbers for charged HL black holes in different ensembles}

The solution of the charged black hole in the HL gravity theory with an electromagnetic field was gotten when $s=1/2$ and $\lambda = 1$ \cite{CCO1}. It is also given by the metric (\ref{2.4}). Now the functions $N(x)= 1$ and $f(x)=k+\frac{x^2}{1-\epsilon^2}- \frac{\sqrt{\epsilon^2 x^4 + (1-\epsilon^2)(c_0 x- q^2/2)}}{1-\epsilon^2}$. Taking the limit $\epsilon \to 1$, the solution becomes $f(x)=k+ \frac{x^2}{2}- \frac{c_0}{2x} +\frac{q^2}{4x^2}$, which is the AdS Reissner-Nordstr\"om black hole solution. Our interest is focused on the solution with $\epsilon^2 = 0$. Taking $\epsilon^2 = 0$, the function is

\begin{eqnarray}
	f(x)=k+x^2-\sqrt{c_0 x-\frac{q^2}{2}},
\label{3.1}
\end{eqnarray}

\noindent where  $x=\sqrt{-\Lambda} r$, $q$ and $c_0$ are integral constants. $c_0$ can be expressed as $c_0=\frac{2k^2+q^2+4k x_{+}^2+2x_{+}^4}{2x_{+}}$, and $x_+ =\sqrt{-\Lambda} r_h$ is determined by $f(x_+)=0$, where $r_h$ is the horizon radius. The black hole's mass and charge are $M=\frac{\kappa^2 \mu^2 \Omega_k \sqrt{-\Lambda} c_0 }{16}$ and $Q=\frac{\kappa^2 \mu^2 \Omega_k \sqrt{-\Lambda} q }{16}$, respectively. We also use speed of light and Newton's constant to rewrite the mass and charge as follows,	

\begin{eqnarray}
	M=\frac{c^3 \Omega_{k} c_0}{16 \pi  G \sqrt{-\Lambda}}, \quad Q=\frac{c^3 \Omega_{k} q}{16 \pi  G \sqrt{-\Lambda}}.
\label{3.2}
\end{eqnarray}	

\noindent The Hawking temperature, entropy and electromagnetic potential are 	

\begin{eqnarray}
	T &=& \sqrt{-\Lambda}\frac{6x_+^4 +4k^2x_+^2 -2k^2-q^2}{16 \pi  k x_+ +16 \pi  x_+^3},\nonumber\\
    S &=& -\frac{c^3 \Omega_k }{4 G \Lambda}\left(2k \ln x_+ + x_+^2 \right) +S_0, \nonumber \\
    \Phi &=& \frac{q}{x_+} + \Phi_0,
\label{3.3}
\end{eqnarray}	

\noindent respectively, where $S_0$ and $\Phi_0$ are constants. In the following, we study the topological numbers for this black hole in canonical and grand canonical ensembles, respectively.

\subsection{Topological numbers in canonical ensemble}

For a canonical ensemble, there is only an exchange of energy between the system and the external environment, and the system's temperature, volume and particle number remain be unchanged. We use the definition of the generalized free energy and get  	

\begin{eqnarray}
	 \mathcal{F}= \frac{c^3 \Omega_{k} c_0}{16 \pi  G \sqrt{-\Lambda}} +\frac{c^3 \Omega_k \left(2k \ln x_+ + x_+^2 \right) +4S_0}{4 G \Lambda \tau}.
\label{3.4}
\end{eqnarray}	

\noindent According to the definition of the vector $\Phi$, its components are 	

\begin{eqnarray}
	\Phi^{r_h}&=&\frac{c^3 \Omega_k  \left(2 k^2+q^2-4 k x_+^2-6 x_+^4\right)}{32 \pi  G x_+^2}-\frac{c^3 \Omega_k  \left(x_+^2 + k\right)}{2 G \tau \sqrt{-\Lambda}x_+},	\nonumber\\
	\Phi ^{\Theta} &=& -\cot\Theta \csc\Theta.
\label{3.5}
\end{eqnarray}	

\noindent We let $\Phi^{r_h} =0$, and get the relation between $\tau$ and $r_h$, which is

\begin{eqnarray}
	\tau=\frac{16 \pi (x_+^3 + kx_+)}{\sqrt{-\Lambda}(6x_+^4 + 4kx_+^2 -2k^2 -q^2)}.
\label{3.6}
\end{eqnarray}

\begin{figure}[H]
	\centering
	\begin{minipage}[t]{0.48\textwidth}
		\centering
		\includegraphics[width=0.6\linewidth]{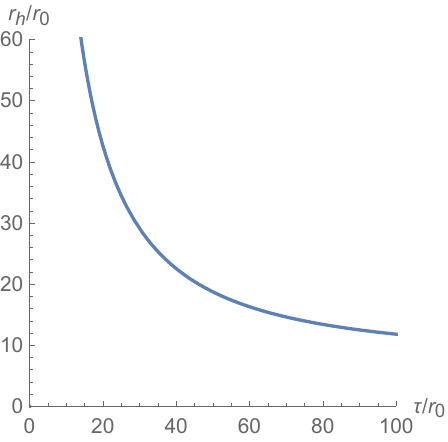}
	\end{minipage}
	\begin{minipage}[t]{0.3\textwidth}
		\centering
		\includegraphics[width=1\linewidth]{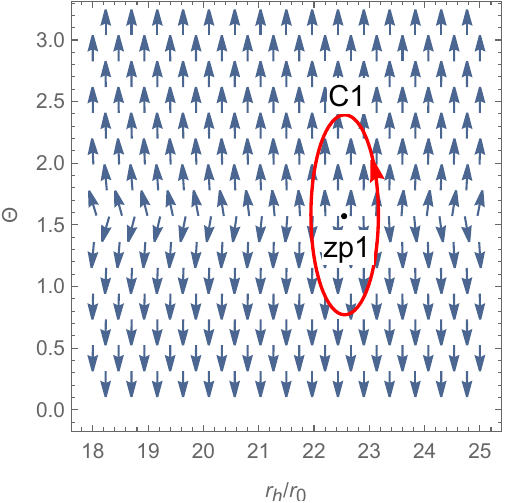}
	\end{minipage}
\caption{Topological properties of the charged HL black hole in the canonical ensemble, where $k = 1$ and $\Lambda r_0^2 = -0.01$. Zero points of the vector $\phi^{r_h}$ in the plane $r_h - \tau$ are plotted in the left picture. The unit vector field $n$ on a portion of the plane $\Theta - r_h $ at $\tau /r_0=40.00 $ is plotted in the right picture. The zero point is at $(r_h/r_0 ,\Theta)$= ($22.54, \pi/2$).}			
\label{fig:3.1.1}
\end{figure}

\begin{figure}[H]
	\centering
	\begin{minipage}[t]{0.48\textwidth}
		\centering
		\includegraphics[width=0.6\linewidth]{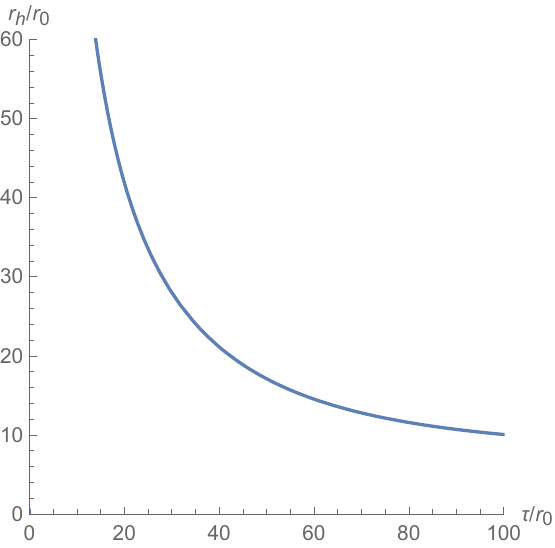}
	\end{minipage}
	\begin{minipage}[t]{0.3\textwidth}
		\centering
		\includegraphics[width=1\linewidth]{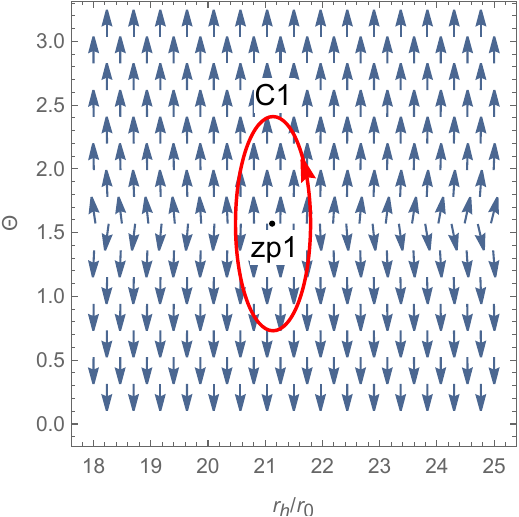}
	\end{minipage}
\caption{Topological properties of the charged HL black hole in the canonical ensemble, where $k = 0$ and $\Lambda r_0^2 = -0.01$. Zero points of the vector $\phi^{r_h}$ in the plane $r_h - \tau$ are plotted in the left picture. The unit vector field $n$ on a portion of the plane $\Theta - r_h $ at $\tau /r_0=40.00 $ is plotted in the right picture. The zero point is at $(r_h/r_0 ,\Theta)$= ($21.12, \pi/2$).}			
\label{fig:3.1.2}
\end{figure}

\begin{figure}[H]
	\centering
	\begin{minipage}[t]{0.48\textwidth}
		\centering
		\includegraphics[width=0.6\linewidth]{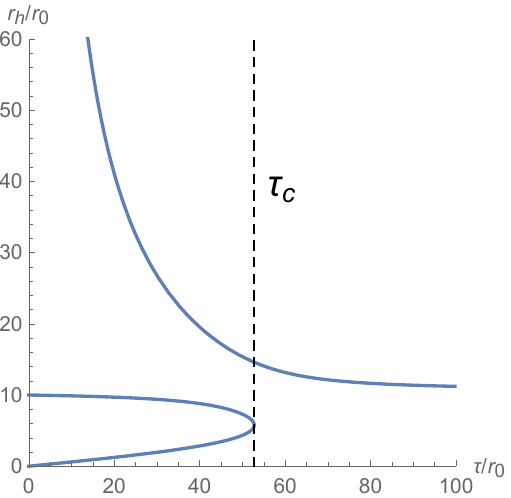}
	\end{minipage}
	\begin{minipage}[t]{0.3\textwidth}
		\centering
		\includegraphics[width=1\linewidth]{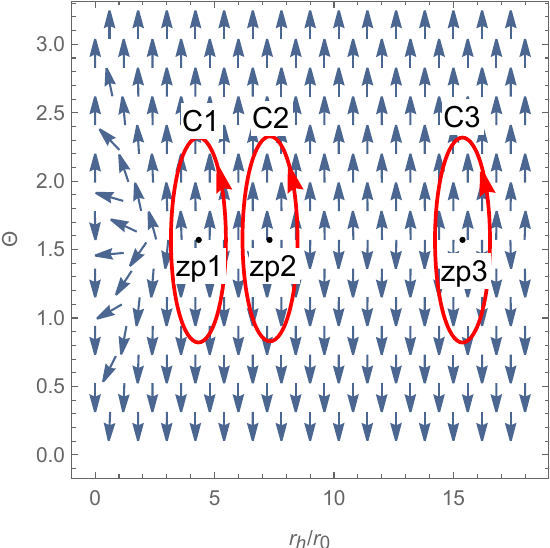}
	\end{minipage}
\caption{Topological properties of the charged HL black hole in the canonical ensemble, where $k = -1$ and $\Lambda r_0^2 = -0.01$. Zero points of the vector $\phi^{r_h}$ in the plane $r_h - \tau$ are plotted in the left picture. The unit vector field $n$ on a portion of the plane $\Theta - r_h $ at $\tau /r_0=50.00 $ is plotted in the right picture. The zero point is at $(r_h/r_0 ,\Theta)$= ($4.34, \pi/2$), ($7.30, \pi/2$) and ($15.38, \pi/2$), respectively.}			
\label{fig:3.1.3}
\end{figure}

In this section, we also order $c=G=\Omega_{k} =1$ and $q= 1$ to plot Figure (\ref{fig:3.1.1}) - (\ref{fig:3.1.3}) and to describe its topological properties. In the figures, we can get the winding numbers from the change of $r_h$ with $\tau$ which reflects the local properties, and the topological numbers which is the sum of the winding numbers.

In Figure (\ref{fig:3.1.1}), the radius of the event horizon monotonically decreases with the increase of $\tau$'s value, which shows that the black hole is stable for any $\tau$'s value and the winding number is $1$. Thus the topological number for the black hole with the spherical horizon is $1$. When $\tau/r_0 =40.00$, the unit vector field $n$ is plotted in the right picture of the figure. There is only one zero point which is at ($22.54, \pi/2$), which also shows the topological number is 1. In \cite{WLM}, the authors have found that the topological number for the four-dimensional RN AdS black hole is 1, therefore, both this black hole and the RN AdS black hole belong to a class with a topological number of 1.

It is also easy to find from Figure (\ref{fig:3.1.2}) that the topological number of the black hole with the flat event horizon is $1$. The zero point is at ($21.12, \pi/2$) in this figure.

Unlike the previous two figures, there are two curves present in Figure (\ref{fig:3.1.3}). The upper curve represents a monotonic decrease of the horizon radius with the increase of $\tau$'s value, which leads to the winding number 1. In the lower left curve, an annihilation point divides the black hole into stable and unstable regions, and their winding numbers are $1$ and $-1$, respectively. We used the approach in \cite{WU6} to calculate the topological number for the black hole with the hyperbolic horizon by considering the combination effect of these two curves. Therefore, the topological number is $1$. The zero points are $(r_h/r_0 ,\Theta)$= ($4.38, \pi/2$), ($7.30, \pi/2$) and ($15.38, \pi/2$), respectively.

\subsection{Topological numbers in grand canonical ensemble}

In the grand canonical ensemble, the system can exchange the energy and charge with the outside, and its temperature, volume and chemical potential remain be unchanged. Now the generalized free energy is defined by 	

\begin{eqnarray}
\mathcal{F}&=& E-Q\Phi -\frac{S}{\tau} \nonumber\\
 &=&  \frac{c^3  \Omega_{k}c_0 }{16\pi  G \sqrt{-\Lambda}} -\frac{c^3 \Omega_k q}{16 \pi  G \sqrt{-\Lambda }}\left(\frac{q}{x_+}+\Phi_0\right)+\frac{c^3\Omega_k \left(2k \ln x_+ + x_+^2 \right) +4S_0}{4 G \Lambda \tau}.
\label{3.7}
\end{eqnarray}	

\noindent We use the definition of the vector $\Phi$, and get its components, 	

\begin{eqnarray}
	\Phi^{r_h}&=&\frac{c^3 \Omega_k  \left(2 k^2-q^2-4 k x_+^2-6 x_+^4\right)}{32 \pi  G x_+^2}-\frac{c^3 \Omega_k  \left(x_+^2 + k\right)}{2 G \tau \sqrt{-\Lambda}x_+},	\nonumber\\
	\Phi ^{\Theta} &=& -\cot\Theta \csc\Theta.
\label{3.8}
\end{eqnarray}	

\noindent Solving $\Phi^{r_h} = 0$ yields the relation

\begin{eqnarray}
	\tau=\frac{16 \pi (x_+^3 + kx_+)}{\sqrt{-\Lambda}(6x_+^4 + 4kx_+^2 -2k^2 +q^2)}.
\label{3.9}
\end{eqnarray}

\begin{figure}[H]
	\centering
	\begin{minipage}[t]{0.48\textwidth}
		\centering
		\includegraphics[width=0.6\linewidth]{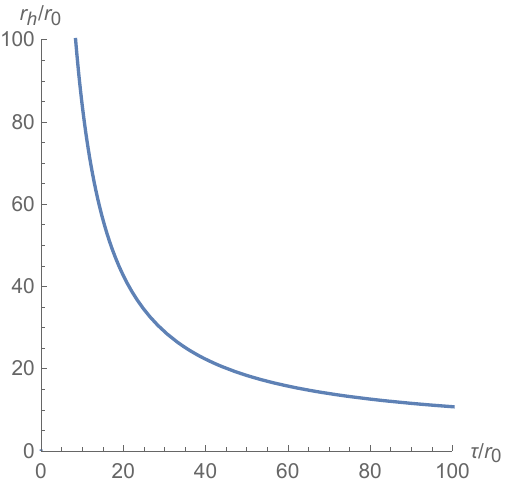}
	\end{minipage}
	\begin{minipage}[t]{0.3\textwidth}
		\centering
		\includegraphics[width=1\linewidth]{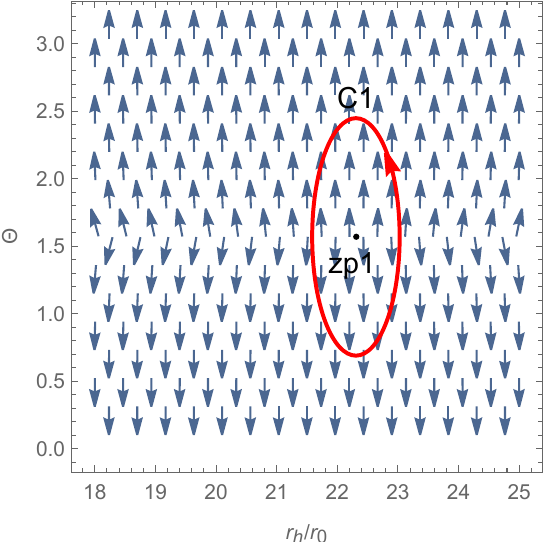}
	\end{minipage}
	\caption{Topological properties of the charged HL black hole in the grand canonical ensemble, where $k = 1$ and $\Lambda r_0^2 = -0.01$. Zero points of the vector $\phi^{r_h}$ in the plane $r_h - \tau$ are plotted in the left picture. The unit vector field $n$ on a portion of the plane $\Theta - r_h $ at $\tau /r_0=40.00 $ is plotted in the right picture. The zero point is at $(r_h/r_0 ,\Theta)$= ($22.31, \pi/2$).}			
	\label{fig:3.2.1}
\end{figure}

\begin{figure}[H]
	\centering
	\begin{minipage}[t]{0.48\textwidth}
		\centering
		\includegraphics[width=0.6\linewidth]{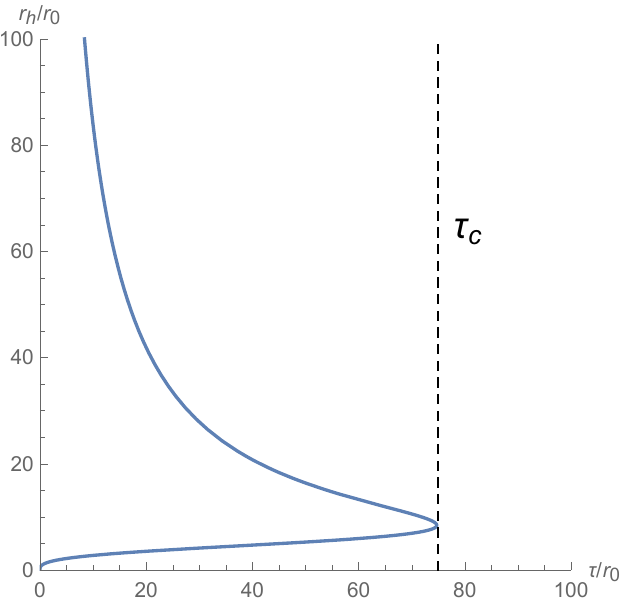}
	\end{minipage}
	\begin{minipage}[t]{0.3\textwidth}
		\centering
		\includegraphics[width=1\linewidth]{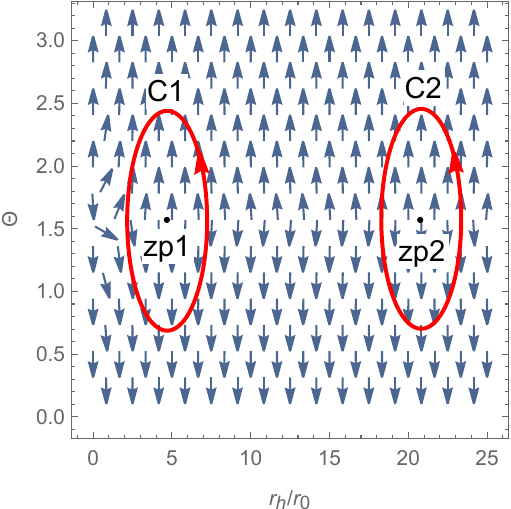}
	\end{minipage}
	\caption{Topological properties of the charged HL black hole in the grand canonical ensemble, where $k = 0$ and $\Lambda r_0^2 = -0.01$. Zero points of the vector $\phi^{r_h}$ in the plane $r_h - \tau$ are plotted in the left picture. The unit vector field $n$ on a portion of the plane $\Theta - r_h $ at $\tau /r_0=40.00 $ is plotted in the right picture. The zero points are at $(r_h/r_0 ,\Theta)$= ($4.68, \pi/2$) and ($20.76, \pi/2$), respectively.}			
	\label{fig:3.2.2}
\end{figure}

\begin{figure}[H]
	\centering
	\begin{minipage}[t]{0.48\textwidth}
		\centering
		\includegraphics[width=0.6\linewidth]{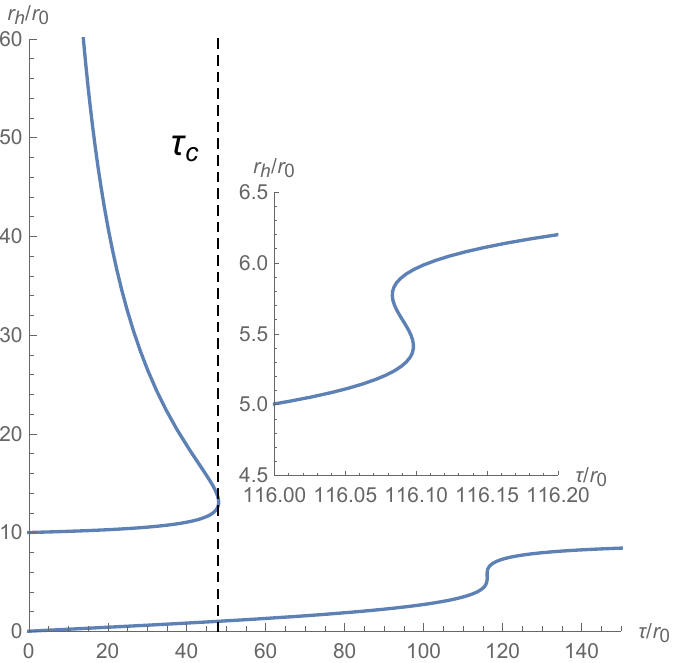}
	\end{minipage}
	\begin{minipage}[t]{0.3\textwidth}
		\centering
		\includegraphics[width=1\linewidth]{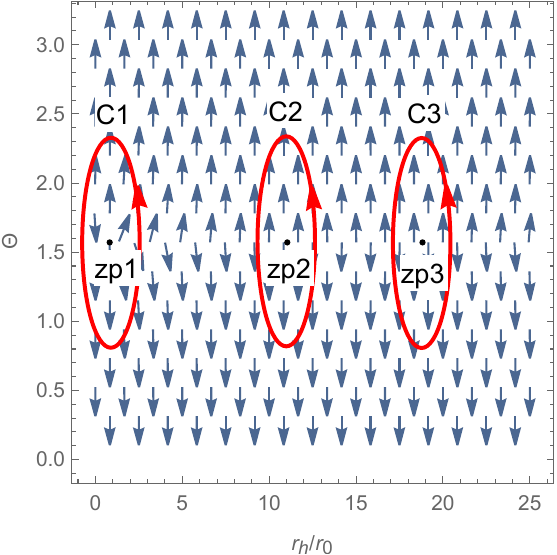}
	\end{minipage}
   	\caption{Topological properties of the charged HL black hole in the grand canonical ensemble, where $k = -1$ and $\Lambda r_0^2 = -0.01$. Zero points of the vector $\phi^{r_h}$ in the plane $r_h - \tau$ are plotted in the left picture. The unit vector field $n$ on a portion of the plane $\Theta - r_h $ at $\tau /r_0=40.00 $ is plotted in the right picture. The zero point is at $(r_h/r_0 ,\Theta)$= ($0.82, \pi/2$), ($11.04, \pi/2$) and ($18.83, \pi/2$), respectively.}			
	\label{fig:3.2.3}
\end{figure}

We use Eqs. (\ref{3.8}) and (\ref{3.9}) and plot Figure (\ref{fig:3.2.1}) - (\ref{fig:3.2.3}) to describe the topological properties. Clearly, the topological number for the black hole with the spherical horizon in Figure (\ref{fig:3.2.1}) is $1$, which is same as that in the canonical ensemble.

In Figure (\ref{fig:3.2.2}), an annihilation point divides the black hole into the stable and unstable regions. In the stable region, the horizon radius monotonically decreases with the increases of $\tau$'s value. In the unstable region, the horizon radius monotonically increases with the increases of $\tau$'s value. Its topological number is $0$. The zero points are at $(r_h/r_0 ,\Theta)$= ($4.68, \pi/2$) and ($20.76, \pi/2$), respectively.

There are two separate curves in Figure (\ref{fig:3.2.3}). The upper left curve has a stable black hole region and an unstable black hole region, with a winding number of 1 and -1, respectively. For another curve, the black hole has a phase transition near an inverse temperature of $\tau/r_0 = 116$, and the $\tau$'s range and the change of the event horizon are not large during the phase transition. The winding numbers reflected on this curve are -1, 1, and -1, respectively. Therefore, the topological number is -1. The right picture shows three zero points, which were obtained at $\tau/r_0= 40.00$. The winding numbers corresponding to these three zeros are -1, 1, and -1, respectively. Therefore, the sum of the three also yields a topological number of -1. It should be noted that when we take $\tau/r_0 = 40.00$, we can only obtain three zero points. When $\tau/r_0 = 116.00$, three other zero points can be obtained, which are not shown in the figure. Considering the number of windings in both cases, we can also obtain a topological number of -1.

Compared with the results in the canonical ensemble, it is not difficult to find that the black hole with the spherical horizon has the same topological number in the canonical and grand canonical ensembles, while the black holes with the flat and hyperbolic horizons have different topological numbers in the canonical and grand canonical ensembles. Therefore, the topological numbers for the black holes with the flat and hyperbolic horizons are ensemble dependent.

\section{Conclusion and discussion}

In this work, we studied the topological numbers for the uncharged and charged black holes in the HL gravity. The influence of the dynamical coupling constant $\lambda$ on the topological numbers for the uncharged black holes has been extensively discussed. For the charged black holes, their topological numbers in the canonical and grand canonical ensembles were studied. The numbers for these black holes obtained in the work are listed in Table \ref{t.1.1} and \ref{t.1.2}.

For the uncharged black holes with the spherical and flat horizons, their topological numbers are the same and independent on the value of the coupling constant. For the uncharged black hole with the hyperbolic horizon, different values of the coupling constant result in different topological numbers, which indicates that the topological number for this black hole is parameter dependent. This coupling constant plays an important role in the topological class of black holes. We have also studied the topological numbers for the charged black holes in the canonical and grand canonical ensembles. In these two ensembles, the topological numbers for the charged black hole with the spherical horizon are the same. While the black holes with the flat and hyperbolic horizons have different topological numbers in these two ensembles. This shows that the last two black holes are ensemble dependent.

On the other hand, according to topological classification, the charged black hole with a spherical horizon and the uncharged black hole belong to the same class because they have the same topological number. Due to the fact that the uncharged and charged black holes with flat and hyperbolic horizons are parameter dependent and ensemble dependent, respectively, it is necessary to consider the parameters' values and ensemble's selection when classifying them. Furthermore, our work is limited to the influence of the dynamic coupling constant on the topological numbers for the static HL black holes. Whether this constant also has an important influence on the number for rotating HL black holes needs to be further confirmed in future work.

\begin{table}[H]
   \centering
   \begin{tabular}{|c|c|c|}
   	\hline
   	Black hole solutions &TNs \\
   	\hline
   	Uncharged BH with the spherical horizon & 1 \\
   	\hline
   	Uncharged BH with the flat horizon & 1\\
   	\hline
   	Uncharged BH with the hyperbolic horizon&  1, 0 or -1\\
   	\hline
   \end{tabular}
   \caption{Topological numbers of uncharged HL black holes. BH is the abbreviation for the HL black hole, and TNs is the abbreviation for the topological numbers.}
   \label{t.1.1}
\end{table}

\begin{table}[H]
   \centering
   \begin{tabular}{|c|c|c|}
   	\hline
   	Black hole solutions &TNs in CE &TNs in GCE\\
   	\hline
   	Charged BH with the spherical horizon & 1  & 1 \\
   	\hline
   	Charged BH with the flat horizon & 1 & 0 \\
   	\hline
   	Charged BH with the hyperbolic horizon& 1 & -1  \\
   	\hline
   \end{tabular}
   \caption{Topological numbers of charged HL black holes in different ensembles. CE is the abbreviation for the canonical ensemble, and GCE is the abbreviation for the grand canonical ensemble.}
   \label{t.1.2}
\end{table}

\end{document}